\documentclass[twocolumn,superscriptaddress,aps,showpacs,amsmath,footinbib,bibnotes]{revtex4-1}
\pdfoutput=1
\usepackage{graphicx}
\usepackage{amssymb,amsmath}
\usepackage{textcomp} 
\usepackage[bold]{hhtensor}
\usepackage{natbib}
\usepackage{lipsum}
\usepackage{bm}
\usepackage{hyperref}
\usepackage{multirow}
\usepackage{float}
\usepackage{amsmath}
\usepackage{amssymb}
\usepackage{stmaryrd}
\usepackage{mathrsfs}
\usepackage[dvipsnames]{xcolor}
\hypersetup{
	colorlinks,
	linkcolor={red!80!black},
	citecolor={Blue},
	urlcolor={MidnightBlue}
}

\begin{document}

\author{Cheng Wang}
\thanks{These authors contributed equally to this work}
\affiliation{John A. Paulson School of Engineering and Applied Sciences, Harvard University, Cambridge, Massachusetts 02138, USA}
\affiliation{Department of Electronic Engineering, City University of Hong Kong, Kowloon, Hong Kong, China}
\author{Mian Zhang}
\thanks{These authors contributed equally to this work}
\affiliation{John A. Paulson School of Engineering and Applied Sciences, Harvard University, Cambridge, Massachusetts 02138, USA}
\author{Rongrong Zhu}
\affiliation{John A. Paulson School of Engineering and Applied Sciences, Harvard University, Cambridge, Massachusetts 02138, USA}
\affiliation{The Electromagnetics Academy at Zhejiang University, College of Information Science and Electronic Engineering, Zhejiang University, Hangzhou 310027, China}
\author{Han Hu}
\affiliation{John A. Paulson School of Engineering and Applied Sciences, Harvard University, Cambridge, Massachusetts 02138, USA}
\affiliation{College of Optical Science and Engineering, Zhejiang University, Hangzhou, 310027, China}
\author{Marko Loncar}\email{loncar@seas.harvard.edu}
\affiliation{John A. Paulson School of Engineering and Applied Sciences, Harvard University, Cambridge, Massachusetts 02138, USA}

\date{\today}
\title{Monolithic photonic circuits for Kerr frequency comb generation, filtering and modulation}

\begin{abstract}
Microresonator Kerr frequency combs, which rely on third-order nonlinearity ($\chi^{(3)}$), are of great interest for a wide range of applications including optical clocks, pulse shaping, spectroscopy, telecommunications, light detection and ranging (LiDAR) and quantum information processing. Many of these applications require further spectral and temporal control of the generated frequency comb signal, which is typically accomplished using additional photonic elements with strong second-order nonlinearity ($\chi^{(2)}$). To date these functionalities have largely been implemented as discrete off-chip components due to material limitations, which come at the expense of extra system complexity and increased optical losses. Here we demonstrate the generation, filtering and electro-optic modulation of a frequency comb on a single monolithic integrated chip, using a thin-film lithium niobate (LN) photonic platform that simultaneously possesses large $\chi^{(2)}$ and $\chi^{(3)}$ nonlinearities and low optical losses. We generate broadband Kerr frequency combs using a dispersion-engineered high quality factor LN microresonator, select a single comb line using an electrically programmable add-drop filter, and modulate the intensity of the selected line. Our results pave the way towards monolithic integrated frequency comb solutions for spectroscopy data communication, ranging and quantum photonics.
\end{abstract}

\maketitle

Many of the microresonator frequency comb applications require not only the comb generator itself, but also additional photonic components such as fast switches, modulators and/or nonlinear wavelength converters, which rely on strong $\chi^{(2)}$ nonlinearity \cite{papp_microresonator_2014,ferdous_spectral_2011,marin-palomo_microresonator-based_2017,reimer_generation_2016}. Chip-scale microresonator Kerr frequency combs have been realized in many material platforms \cite{kippenberg_microresonator-based_2011}, including silica (SiO$_2$) \cite{papp_microresonator_2014,delhaye_frequency_2009,suh_soliton_2018,kippenberg_kerr-nonlinearity_2004} silicon nitride (Si$_3$N$_4$) \cite{papp_microresonator_2014,dutt_-chip_2018,trocha_ultrafast_2018,marin-palomo_microresonator-based_2017,reimer_generation_2016,okawachi_octave-spanning_2011} silicon (Si) \cite{griffith_silicon-chip_2015}, crystalline fluorides \cite{herr_temporal_2014}, diamond \cite{hausmann_diamond_2014}, aluminium nitride (AlN) \cite{jung_green_2014}, and aluminium-gallium arsenide (AlGaAs) \cite{pu_efficient_2016}. While most of these materials possess large $\chi^{(3)}$ nonlinearity and low optical loss, required for Kerr comb generation, they usually have small or zero $\chi^{(2)}$ nonlinearity and therefore are not suitable for on-chip integration of the aforementioned $\chi^{(2)}$ components. Carrier-injection based Si devices can be electrically modulated at high speeds, but exhibits much higher optical losses than their intrinsic Si counterparts \cite{reed_silicon_2010}. AlGaAs possesses high $\chi^{(2)}$ nonlinearity for second harmonic generation, but much weaker electro-optic effect ($r_{41} = 1.5 \times 10^{-12}$ m/V) \cite{nikogosyan_nonlinear_2005}. As a result, off-chip components are typically used for achieving these complex electro-optic and nonlinear optic functionalities \cite{papp_microresonator_2014,marin-palomo_microresonator-based_2017,ferdous_spectral_2011,reimer_generation_2016} and to date on-chip manipulation of the generated combs has been limited to slow thermal effects \cite{miller_tunable_2015} or high-voltage electrical signals \cite{jung_electrical_2014}. While heterogeneous integration of photonic chips with different functionalities has been proposed to circumvent this problem \cite{spencer_optical-frequency_2018}, this approach increases the complexity and cost of the system, and requires scalable and low-loss optical links between chips. 

\begin{figure*}
	\centering
	\includegraphics[angle=0,width=\textwidth]{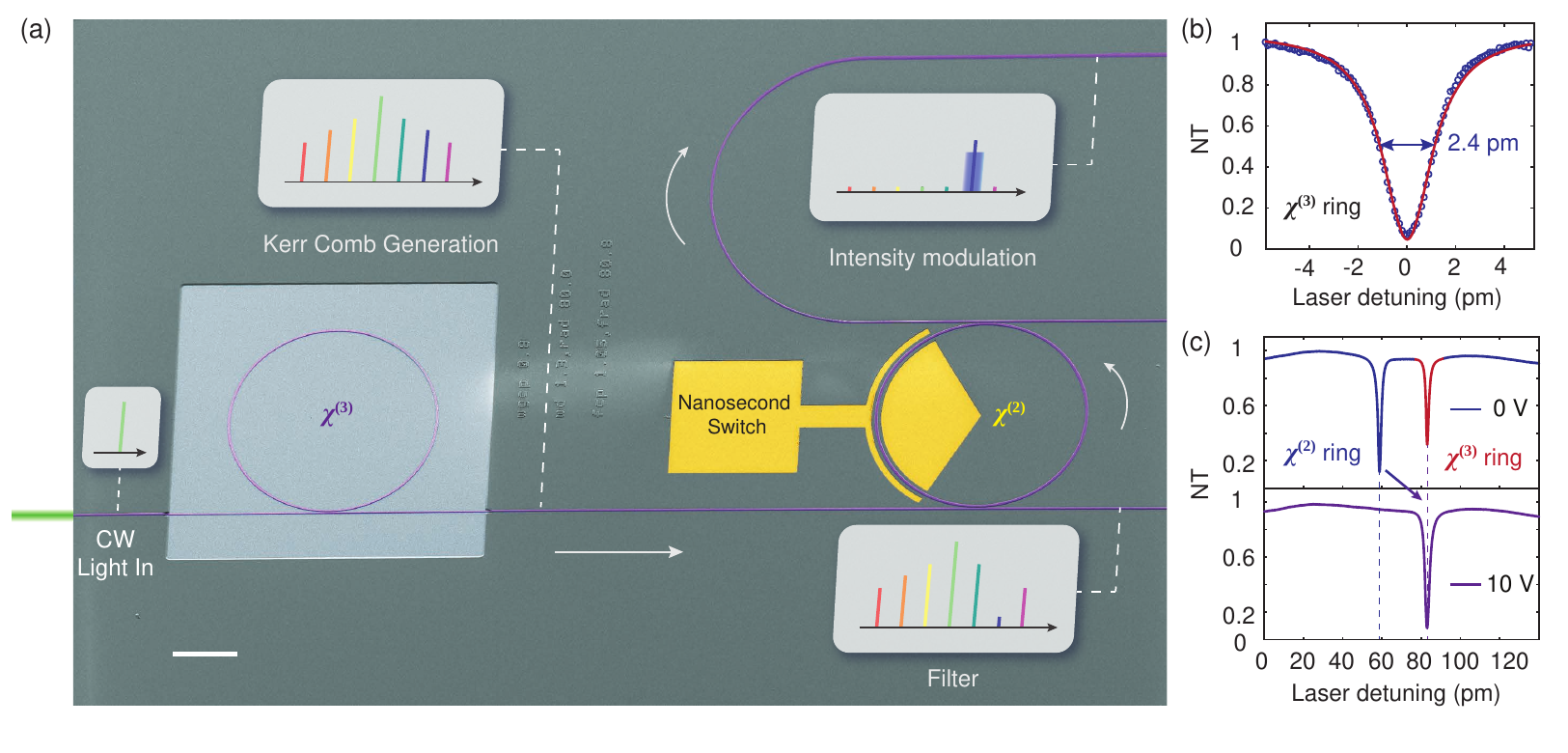}
	
	\caption{\label{fig1}\textbf{Monolithic integrated photonic circuit for frequency comb generation and manipulation.} \textbf{(a)} A false-colour scanning electron microscope (SEM) image showing a fabricated lithium-niobate nanophotonic circuit that consists of a microresonator frequency comb generator ($\chi^{(3)}$) and an electro-optically tuneable add-drop filter ($\chi^{(2)}$). The comb generation area is air cladded to achieve anomalous dispersion, whereas the rest of the chip is cladded in SiO$_2$. Continuous-wave (CW) pump light first passes through the dispersion-engineered microring resonator to generate a frequency comb. The generated frequency comb is then filtered by an add-drop microring filter. At the drop port of the filter, a single target comb line is selected by applying an external bias voltage on the integrated electrodes to align the filter passband with the comb line. Finally, the selected comb line can be modulated at high speeds via the $\chi^{(2)}$ effect. \textbf{(b)} Optical transmission spectrum of the $\chi^{(3)}$ microring resonator. The measured loaded (intrinsic) quality ($Q$) factor of transverse-electric (TE) polarized mode is $6.6 \times 10^5 (1.1 \times 10^6)$. \textbf{(c)} Transmission spectra at the through port when different DC bias voltages are applied. At zero bias (blue curve), the comb resonance has a 24-pm mismatch with the filter resonance. Applying a bias of 10 V can align the two resonances (red curve), showing a measured electrical tuning efficiency of 2.4 pm/V. Scale bar: 50 $\mu$m.}
\end{figure*}

Here we address this challenge by the monolithic integration of lithium niobate (LN, LiNbO$_3$) nanophotonic waveguides, microring Kerr comb generators, filters and electro-optic modulators on the same chip. LN is a material that simultaneously possess large $\chi^{(3)}$ (Kerr) ($1.6 \times 10^{-21} $m$^2$/V$^2$) and $\chi^{(2)}$ ($r_{33} = 3 \times 10^{-11}$ m/V) nonlinearities \cite{nikogosyan_nonlinear_2005,desalvo_infrared_1996}. Specifically, the $\chi^{(3)}$ nonlinearity enables the generation of Kerr frequency combs, whereas the $\chi^{(2)}$ nonlinearity (electro-optic effect) is used to manipulate the generated comb by an external electrical field (Fig. \ref{fig1}). 

\begin{figure*}
	\centering
	\includegraphics[angle=0,width=\textwidth]{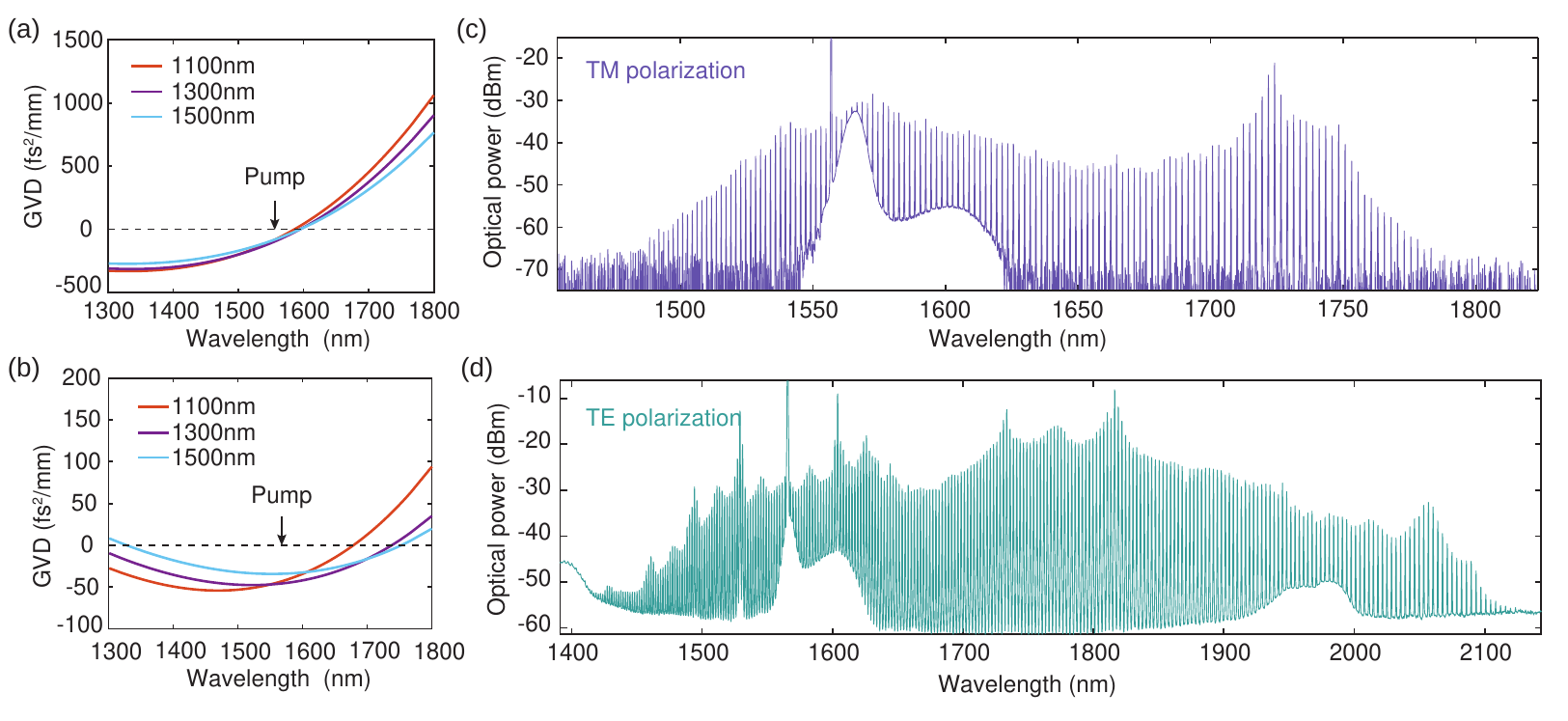}
	
	\caption{\label{fig2}\textbf{Broadband frequency comb generation.} \textbf{(a-b)} Numerically simulated group-velocity dispersions (GVD) at telecom wavelengths for LN waveguides with different top widths. Anomalous dispersions (GVD $<$ 0) can be achieved for both transverse-magnetic (TM) (a) and transverse-electric (TE) (b) modes. \textbf{(c-d)}, Generated frequency comb spectra when the input laser is tuned into resonance with either TM (c) or TE (d) modes, showing comb spans of $\sim$ 300 nm (c) and $\sim$ 700 nm (d), respectively.}
\end{figure*}

In order for the $\chi^{(3)}$ optical parametric oscillation (OPO) process to take place, a microresonator with a high quality ($Q$) factor and anomalous dispersion is needed. The former ensures that the four-wave mixing (FWM) process could cascade and overcome the optical losses of the microresonator, and the latter compensates for the nonlinear responses of the strong pump, i.e. self-phase modulation (SPM) and cross-phase modulation (XPM) \cite{kippenberg_kerr-nonlinearity_2004}. While ultra-high-$Q$ ($\sim 10^8$) LN whispering-gallery-mode resonators have been demonstrated using mechanical polishing methods \cite{ilchenko_nonlinear_2004}, their dispersion properties are predetermined by the bulk material properties and cannot be engineered. In contrast, our integrated approach relies on an ultralow-loss micro-structured LN photonic platform that offers dispersion engineering capability. Our platform utilizes a single-crystal LN film with sub-micron thickness bonded on top of a SiO$_2$ substrate \cite{wang_nanophotonic_2018,zhang_monolithic_2017,he_dispersion-engineered_2018,wang_integrated_2014,wolf_cascaded_2017,wang_high-q_2015,liang_high-quality_2017}. By lithography and dry etching of the thin LN film, microresonators that have $Q$ factors up to $10^7$ \cite{zhang_monolithic_2017}, and that allow dispersion engineering \cite{he_dispersion-engineered_2018}, can be realized. Using an x-cut LN thin-film wafer, we achieve anomalous dispersion in the telecom wavelength range for both the transverse-electric (TE) and transverse-magnetic (TM) polarizations by carefully engineering the waveguide width and thickness (Fig. \ref{fig2}a-b). Our dispersion-engineered microring resonator feature a loaded (intrinsic) $Q$ factor of $6.6 \times 10^5 (1.1 \times 10^6)$ for TE polarization, as shown in Fig. \ref{fig1}b, resulting in an estimated OPO pump threshold of $\sim 80$ mW (see Supplementary) \cite{ilchenko_nonlinear_2004}. The loaded (intrinsic) $Q$ factor of the TM mode is $6.0 \times 10^5 (9.2 \times 10^5$). 

For a microring resonator with a radius of 80 $\mu$m and a top width of 1.3 $\mu$m, we observe broadband frequency comb generation for both TE-like and TM-like polarization modes at a pump power of 300 mW in the input bus waveguide, with a comb line spacing of $\sim 2$ nm (250 GHz) (Fig. \ref{fig2}c-d). The measured TM-polarized comb spectrum is $\sim$ 300 nm wide, while the TE-polarized comb spans from 1400 nm to 2100 nm, over two-thirds of an octave. The envelopes of the comb spectra indicate that the generated combs are likely not in a soliton state, i.e. are modulation instability frequency combs \cite{herr_temporal_2014}. Soliton states can potentially be achieved using temporal scanning techniques that have been deployed in other material platforms \cite{herr_temporal_2014}. Importantly, our integrated LN resonators can sustain high optical powers ($\sim 50$ W of circulating power), unlike their bulk/ion-diffused LN counterparts, where the photorefractive effect often causes device instability and/or irreversible damage. In our devices, the photorefractive effect shows quenching behaviour at high pump powers ($>$ 100 mW in the waveguide), similar to what was previously observed \cite{liang_high-quality_2017}. As a result, the thermal bistability effect dominates, allowing us to stably position the laser detuning with respect to cavity resonance. Despite the high circulating power inside our resonators, we do not observe optical damage after many hours of optical pumping.

We achieve the filtering and fast modulation of comb signals by integrating an electrically tuneable add-drop filter with the comb generator on the same chip (Fig. \ref{fig1}). The add-drop filter consists of a LN microring resonator whose free-spectral range (FSR) is designed to be $\sim 1\%$ larger than the comb generator (Supplementary Fig. \ref{figs1}). The slightly detuned FSR utilizes the Vernier effect to enable the selection of a single optical spectral line over a wide optical band. The filter ring is over-coupled to both the add and the drop bus waveguides with the same coupling strength, to ensure a high extinction ratio (on/off ratio). When the input light is on (off) resonance with the filter, the majority of the optical power at the wavelength of interest will be transmitted to the drop (through) port of the filter. Importantly, the microring filter is integrated with metal electrodes positioned close to the ring. This allows for fast and efficient tuning of the filter frequency (Fig. \ref{fig1}c), as well as amplitude modulation of the dropped light, via the electro-optic effect. In order to access the maximum electro-optic coefficient ($r_{33}$), we design the two resonators to operate both in TE modes. The comb ring and the filter ring are cladded with air and SiO$_2$ respectively (Fig. \ref{fig1}a), to ensure that both devices operate in their best configurations (see Supplementary). 

\begin{figure*}[t!]
	\centering
	\includegraphics[angle=0,width=0.7\textwidth]{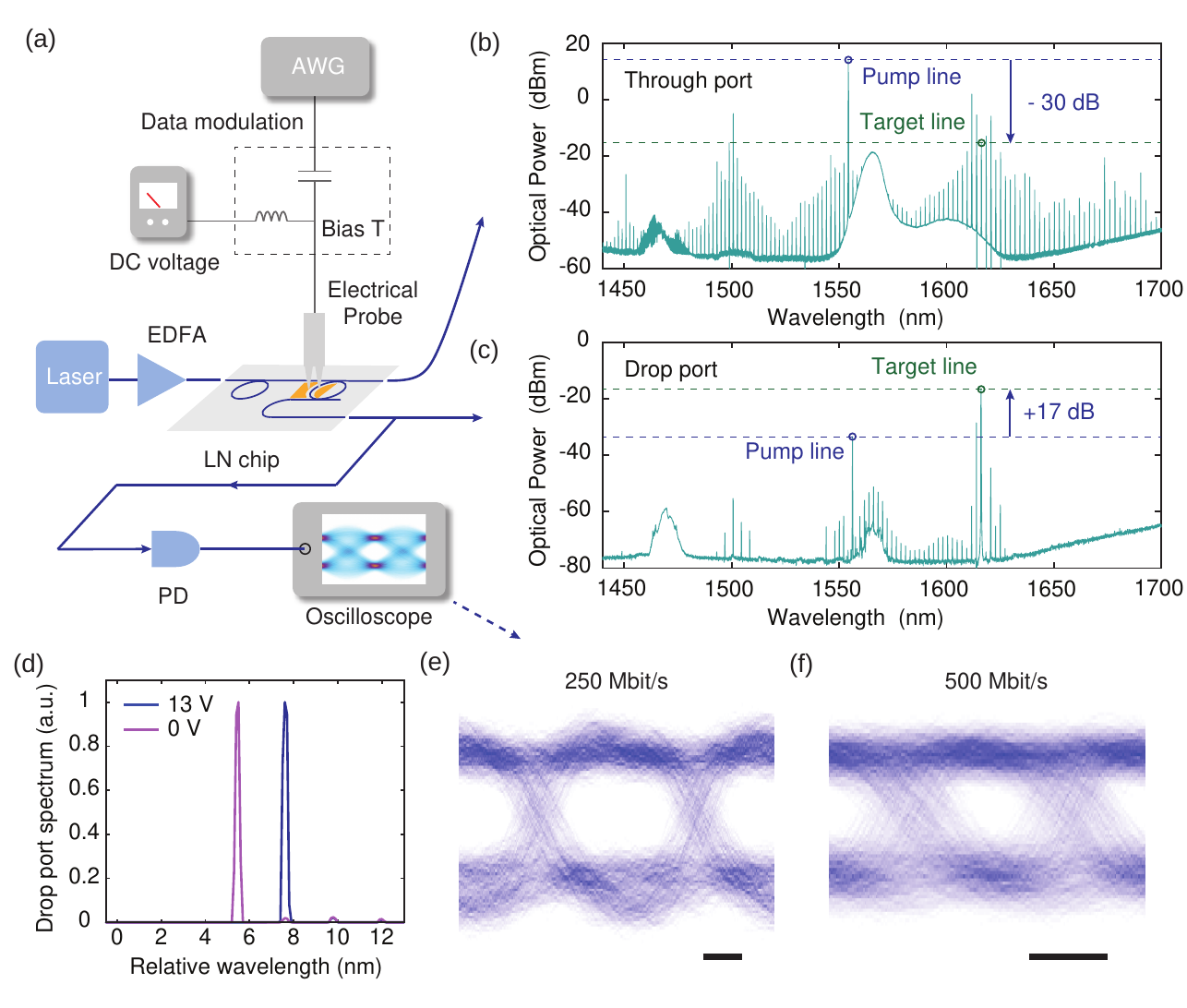}
	
	\caption{\label{fig3}\textbf{On-chip filtering and modulation of a frequency comb.} \textbf{(a)} Simplified characterization setup. \textbf{(b-c)} Measured optical spectra at the through (b) and the drop (c) ports of the filter, picking out a target comb line at $\sim$ 1616 nm. The filter shows 47 dB suppression of the pump light. \textbf{(d)} Zoom-in view of the drop-port output spectra near the target line at different DC bias voltages. Applying a bias voltage of 13 V shifts the target from one comb line to the next one. \textbf{(e-f)}, Applying AC electric signals could modulate the intensity of the selected comb line at 250 Mbit/s (e) and 500 Mbit/s (f). Eye diagrams are measured by sending a random-binary voltage sequence to the filter, and monitoring the real-time output optical power. Open-eye operations can be achieved for both bit rates. Scale bars: 1 ns. AWG, arbitrary waveform generator; EDFA, erbium-doped fiber amplifier; PD, photodetector.}
\end{figure*}

We show efficient filtering of a single comb line and the strong suppression of pump light using the on-chip filter (Fig. \ref{fig3}a-d). We apply a direct-current (DC) bias voltage to align the filter frequency with a target comb line at 1616 nm (Fig. \ref{fig1}c). In this case, the pump frequency at $\sim$ 1556 nm has a 730-pm mismatch with the filter resonance, resulting in an experimentally measured 47 dB rejection of the pump power in the drop port (Fig. \ref{fig3}b-c). The filter also shows 20 dB extinction for the comb lines adjacent to the target line (Fig. \ref{fig3}b-c). The measured filter extinction ratios agree well with theoretical predictions (Supplementary Fig. \ref{figs2}). Different target comb lines can be selected by applying different bias voltages (Fig. \ref{fig3}d). The required additional bias voltage to change the target comb line to the adjacent one is measured to be 13 V (Fig. \ref{fig3}d). This is consistent with the measured electro-optic tuning efficiency of 2.4 pm/V and the FSR difference between the comb resonator and filter resonator of 27 pm. 

We show the selected target comb line at the drop port can also be modulated at high speeds, far exceeding previously demonstrated tuning method based on thermo-optic effects \cite{miller_tunable_2015}. We use an arbitrary waveform generator (AWG) to deliver random-binary voltage sequences to the electrodes of the filter ring, in addition to the DC bias voltage (Fig. \ref{fig3}a). The peak-to-peak modulation voltage in this case is 1.5 V, sufficient to tune the filter passband (3 pm wide) away from the target comb line. At data rates of 250 Mbit/s and 500 Mbit/s, we demonstrate open-eye data operation of the filtered comb line (Fig. \ref{fig3}e-f). The electro-optic bandwidth of our filter/modulator ($\sim 400$ MHz) is currently limited by the photon lifetime of the resonator (0.4 ns). The modulation speed can be dramatically improved (beyond 100 Gbit/s) by integrating a Mach-Zehnder modulator after the tuneable microring filter \cite{zhang_ultra-high_2018}.

In summary, we have demonstrated Kerr comb generation followed by spectral and temporal manipulation of the comb signal, all achieved on the same LN chip. Our platform could lead to a new generation of photonic circuits based on the monolithic integration of frequency comb generators with both passive and active photonic components. Leveraging the giant effective $\chi^{(3)}$ nonlinearity in a quasi-phase-matched $\chi^{(2)}$ waveguide, frequency comb generation with much lower threshold power could potentially be achieved \cite{phillips_supercontinuum_2011}. Directly embedding electro-optic modulation in the comb generator could lead to active mode locking of a Kerr frequency comb. Further integrating the frequency comb source with multiplexer/demultiplexer and ultrafast electro-optic modulators on the same chip could provide compact and low-cost dense-wavelength-division multiplexing (DWDM) solutions for future ultra-broadband optical fiber communication networks \cite{marin-palomo_microresonator-based_2017}. The fast and independent control of the amplitude and phase of each comb line are promising for chip-scale LiDAR systems \cite{trocha_ultrafast_2018,suh_soliton_2018} programmable pulse shaping \cite{ferdous_spectral_2011} and quantum information processing \cite{reimer_generation_2016}. 

The authors thank C. Reimer for valuable discussions. This work is supported in part by National Science Foundation (NSF) (ECCS1609549, ECCS-1740296 E2CDA), by Harvard University Office of Technology Development (Physical Sciences and Engineering Accelerator Award), and DARPA SCOUT program (W31P4Q-15-1-0013). Device fabrication is performed at the Harvard University Center for Nanoscale Systems (CNS), a member of the National Nanotechnology Coordinated Infrastructure Network (NNCI), which is supported by the National Science Foundation under NSF ECCS award no. 1541959.

\bibliography{KerrComb}

\begin{thebibliography}{32}%
\makeatletter
\providecommand \@ifxundefined [1]{%
 \@ifx{#1\undefined}
}%
\providecommand \@ifnum [1]{%
 \ifnum #1\expandafter \@firstoftwo
 \else \expandafter \@secondoftwo
 \fi
}%
\providecommand \@ifx [1]{%
 \ifx #1\expandafter \@firstoftwo
 \else \expandafter \@secondoftwo
 \fi
}%
\providecommand \natexlab [1]{#1}%
\providecommand \enquote  [1]{``#1''}%
\providecommand \bibnamefont  [1]{#1}%
\providecommand \bibfnamefont [1]{#1}%
\providecommand \citenamefont [1]{#1}%
\providecommand \href@noop [0]{\@secondoftwo}%
\providecommand \href [0]{\begingroup \@sanitize@url \@href}%
\providecommand \@href[1]{\@@startlink{#1}\@@href}%
\providecommand \@@href[1]{\endgroup#1\@@endlink}%
\providecommand \@sanitize@url [0]{\catcode `\\12\catcode `\$12\catcode
  `\&12\catcode `\#12\catcode `\^12\catcode `\_12\catcode `\%12\relax}%
\providecommand \@@startlink[1]{}%
\providecommand \@@endlink[0]{}%
\providecommand \url  [0]{\begingroup\@sanitize@url \@url }%
\providecommand \@url [1]{\endgroup\@href {#1}{\urlprefix }}%
\providecommand \urlprefix  [0]{URL }%
\providecommand \Eprint [0]{\href }%
\providecommand \doibase [0]{http://dx.doi.org/}%
\providecommand \selectlanguage [0]{\@gobble}%
\providecommand \bibinfo  [0]{\@secondoftwo}%
\providecommand \bibfield  [0]{\@secondoftwo}%
\providecommand \translation [1]{[#1]}%
\providecommand \BibitemOpen [0]{}%
\providecommand \bibitemStop [0]{}%
\providecommand \bibitemNoStop [0]{.\EOS\space}%
\providecommand \EOS [0]{\spacefactor3000\relax}%
\providecommand \BibitemShut  [1]{\csname bibitem#1\endcsname}%
\let\auto@bib@innerbib\@empty
\bibitem [{\citenamefont {Papp}\ \emph {et~al.}(2014)\citenamefont {Papp},
  \citenamefont {Beha}, \citenamefont {Del’Haye}, \citenamefont {Quinlan},
  \citenamefont {Lee}, \citenamefont {Vahala},\ and\ \citenamefont
  {Diddams}}]{papp_microresonator_2014}%
  \BibitemOpen
  \bibfield  {author} {\bibinfo {author} {\bibfnamefont {S.~B.}\ \bibnamefont
  {Papp}}, \bibinfo {author} {\bibfnamefont {K.}~\bibnamefont {Beha}}, \bibinfo
  {author} {\bibfnamefont {P.}~\bibnamefont {Del’Haye}}, \bibinfo {author}
  {\bibfnamefont {F.}~\bibnamefont {Quinlan}}, \bibinfo {author} {\bibfnamefont
  {H.}~\bibnamefont {Lee}}, \bibinfo {author} {\bibfnamefont {K.~J.}\
  \bibnamefont {Vahala}}, \ and\ \bibinfo {author} {\bibfnamefont {S.~A.}\
  \bibnamefont {Diddams}},\ }\href {\doibase 10.1364/OPTICA.1.000010}
  {\bibfield  {journal} {\bibinfo  {journal} {Optica}\ }\textbf {\bibinfo
  {volume} {1}},\ \bibinfo {pages} {10} (\bibinfo {year} {2014})}\BibitemShut
  {NoStop}%
\bibitem [{\citenamefont {Ferdous}\ \emph {et~al.}(2011)\citenamefont
  {Ferdous}, \citenamefont {Miao}, \citenamefont {Leaird}, \citenamefont
  {Srinivasan}, \citenamefont {Wang}, \citenamefont {Chen}, \citenamefont
  {Varghese},\ and\ \citenamefont {Weiner}}]{ferdous_spectral_2011}%
  \BibitemOpen
  \bibfield  {author} {\bibinfo {author} {\bibfnamefont {F.}~\bibnamefont
  {Ferdous}}, \bibinfo {author} {\bibfnamefont {H.}~\bibnamefont {Miao}},
  \bibinfo {author} {\bibfnamefont {D.~E.}\ \bibnamefont {Leaird}}, \bibinfo
  {author} {\bibfnamefont {K.}~\bibnamefont {Srinivasan}}, \bibinfo {author}
  {\bibfnamefont {J.}~\bibnamefont {Wang}}, \bibinfo {author} {\bibfnamefont
  {L.}~\bibnamefont {Chen}}, \bibinfo {author} {\bibfnamefont {L.~T.}\
  \bibnamefont {Varghese}}, \ and\ \bibinfo {author} {\bibfnamefont {A.~M.}\
  \bibnamefont {Weiner}},\ }\href {\doibase 10.1038/nphoton.2011.255}
  {\bibfield  {journal} {\bibinfo  {journal} {Nature Photonics}\ }\textbf
  {\bibinfo {volume} {5}},\ \bibinfo {pages} {770} (\bibinfo {year}
  {2011})}\BibitemShut {NoStop}%
\bibitem [{\citenamefont {Marin-Palomo}\ \emph {et~al.}(2017)\citenamefont
  {Marin-Palomo}, \citenamefont {Kemal}, \citenamefont {Karpov}, \citenamefont
  {Kordts}, \citenamefont {Pfeifle}, \citenamefont {Pfeiffer}, \citenamefont
  {Trocha}, \citenamefont {Wolf}, \citenamefont {Brasch}, \citenamefont
  {Anderson}, \citenamefont {Rosenberger}, \citenamefont {Vijayan},
  \citenamefont {Freude}, \citenamefont {Kippenberg},\ and\ \citenamefont
  {Koos}}]{marin-palomo_microresonator-based_2017}%
  \BibitemOpen
  \bibfield  {author} {\bibinfo {author} {\bibfnamefont {P.}~\bibnamefont
  {Marin-Palomo}}, \bibinfo {author} {\bibfnamefont {J.~N.}\ \bibnamefont
  {Kemal}}, \bibinfo {author} {\bibfnamefont {M.}~\bibnamefont {Karpov}},
  \bibinfo {author} {\bibfnamefont {A.}~\bibnamefont {Kordts}}, \bibinfo
  {author} {\bibfnamefont {J.}~\bibnamefont {Pfeifle}}, \bibinfo {author}
  {\bibfnamefont {M.~H.~P.}\ \bibnamefont {Pfeiffer}}, \bibinfo {author}
  {\bibfnamefont {P.}~\bibnamefont {Trocha}}, \bibinfo {author} {\bibfnamefont
  {S.}~\bibnamefont {Wolf}}, \bibinfo {author} {\bibfnamefont {V.}~\bibnamefont
  {Brasch}}, \bibinfo {author} {\bibfnamefont {M.~H.}\ \bibnamefont
  {Anderson}}, \bibinfo {author} {\bibfnamefont {R.}~\bibnamefont
  {Rosenberger}}, \bibinfo {author} {\bibfnamefont {K.}~\bibnamefont
  {Vijayan}}, \bibinfo {author} {\bibfnamefont {W.}~\bibnamefont {Freude}},
  \bibinfo {author} {\bibfnamefont {T.~J.}\ \bibnamefont {Kippenberg}}, \ and\
  \bibinfo {author} {\bibfnamefont {C.}~\bibnamefont {Koos}},\ }\href {\doibase
  10.1038/nature22387} {\bibfield  {journal} {\bibinfo  {journal} {Nature}\
  }\textbf {\bibinfo {volume} {546}},\ \bibinfo {pages} {274} (\bibinfo {year}
  {2017})}\BibitemShut {NoStop}%
\bibitem [{\citenamefont {Reimer}\ \emph {et~al.}(2016)\citenamefont {Reimer},
  \citenamefont {Kues}, \citenamefont {Roztocki}, \citenamefont {Wetzel},
  \citenamefont {Grazioso}, \citenamefont {Little}, \citenamefont {Chu},
  \citenamefont {Johnston}, \citenamefont {Bromberg}, \citenamefont {Caspani},
  \citenamefont {Moss},\ and\ \citenamefont
  {Morandotti}}]{reimer_generation_2016}%
  \BibitemOpen
  \bibfield  {author} {\bibinfo {author} {\bibfnamefont {C.}~\bibnamefont
  {Reimer}}, \bibinfo {author} {\bibfnamefont {M.}~\bibnamefont {Kues}},
  \bibinfo {author} {\bibfnamefont {P.}~\bibnamefont {Roztocki}}, \bibinfo
  {author} {\bibfnamefont {B.}~\bibnamefont {Wetzel}}, \bibinfo {author}
  {\bibfnamefont {F.}~\bibnamefont {Grazioso}}, \bibinfo {author}
  {\bibfnamefont {B.~E.}\ \bibnamefont {Little}}, \bibinfo {author}
  {\bibfnamefont {S.~T.}\ \bibnamefont {Chu}}, \bibinfo {author} {\bibfnamefont
  {T.}~\bibnamefont {Johnston}}, \bibinfo {author} {\bibfnamefont
  {Y.}~\bibnamefont {Bromberg}}, \bibinfo {author} {\bibfnamefont
  {L.}~\bibnamefont {Caspani}}, \bibinfo {author} {\bibfnamefont {D.~J.}\
  \bibnamefont {Moss}}, \ and\ \bibinfo {author} {\bibfnamefont
  {R.}~\bibnamefont {Morandotti}},\ }\href {\doibase 10.1126/science.aad8532}
  {\bibfield  {journal} {\bibinfo  {journal} {Science}\ }\textbf {\bibinfo
  {volume} {351}},\ \bibinfo {pages} {1176} (\bibinfo {year}
  {2016})}\BibitemShut {NoStop}%
\bibitem [{\citenamefont {Kippenberg}\ \emph {et~al.}(2011)\citenamefont
  {Kippenberg}, \citenamefont {Holzwarth},\ and\ \citenamefont
  {Diddams}}]{kippenberg_microresonator-based_2011}%
  \BibitemOpen
  \bibfield  {author} {\bibinfo {author} {\bibfnamefont {T.~J.}\ \bibnamefont
  {Kippenberg}}, \bibinfo {author} {\bibfnamefont {R.}~\bibnamefont
  {Holzwarth}}, \ and\ \bibinfo {author} {\bibfnamefont {S.~A.}\ \bibnamefont
  {Diddams}},\ }\href {\doibase 10.1126/science.1193968} {\bibfield  {journal}
  {\bibinfo  {journal} {Science}\ }\textbf {\bibinfo {volume} {332}},\ \bibinfo
  {pages} {555} (\bibinfo {year} {2011})}\BibitemShut {NoStop}%
\bibitem [{\citenamefont {Del'Haye}\ \emph {et~al.}(2009)\citenamefont
  {Del'Haye}, \citenamefont {Arcizet}, \citenamefont {Gorodetsky},
  \citenamefont {Holzwarth},\ and\ \citenamefont
  {Kippenberg}}]{delhaye_frequency_2009}%
  \BibitemOpen
  \bibfield  {author} {\bibinfo {author} {\bibfnamefont {P.}~\bibnamefont
  {Del'Haye}}, \bibinfo {author} {\bibfnamefont {O.}~\bibnamefont {Arcizet}},
  \bibinfo {author} {\bibfnamefont {M.~L.}\ \bibnamefont {Gorodetsky}},
  \bibinfo {author} {\bibfnamefont {R.}~\bibnamefont {Holzwarth}}, \ and\
  \bibinfo {author} {\bibfnamefont {T.~J.}\ \bibnamefont {Kippenberg}},\ }\href
  {\doibase 10.1038/nphoton.2009.138} {\bibfield  {journal} {\bibinfo
  {journal} {Nature Photonics}\ }\textbf {\bibinfo {volume} {3}},\ \bibinfo
  {pages} {529} (\bibinfo {year} {2009})}\BibitemShut {NoStop}%
\bibitem [{\citenamefont {Suh}\ and\ \citenamefont
  {Vahala}(2018)}]{suh_soliton_2018}%
  \BibitemOpen
  \bibfield  {author} {\bibinfo {author} {\bibfnamefont {M.-G.}\ \bibnamefont
  {Suh}}\ and\ \bibinfo {author} {\bibfnamefont {K.~J.}\ \bibnamefont
  {Vahala}},\ }\href {\doibase 10.1126/science.aao1968} {\bibfield  {journal}
  {\bibinfo  {journal} {Science}\ }\textbf {\bibinfo {volume} {359}},\ \bibinfo
  {pages} {884} (\bibinfo {year} {2018})}\BibitemShut {NoStop}%
\bibitem [{\citenamefont {Kippenberg}\ \emph {et~al.}(2004)\citenamefont
  {Kippenberg}, \citenamefont {Spillane},\ and\ \citenamefont
  {Vahala}}]{kippenberg_kerr-nonlinearity_2004}%
  \BibitemOpen
  \bibfield  {author} {\bibinfo {author} {\bibfnamefont {T.~J.}\ \bibnamefont
  {Kippenberg}}, \bibinfo {author} {\bibfnamefont {S.~M.}\ \bibnamefont
  {Spillane}}, \ and\ \bibinfo {author} {\bibfnamefont {K.~J.}\ \bibnamefont
  {Vahala}},\ }\href {\doibase 10.1103/PhysRevLett.93.083904} {\bibfield
  {journal} {\bibinfo  {journal} {Physical Review Letters}\ }\textbf {\bibinfo
  {volume} {93}},\ \bibinfo {pages} {083904} (\bibinfo {year}
  {2004})}\BibitemShut {NoStop}%
\bibitem [{\citenamefont {Dutt}\ \emph {et~al.}(2018)\citenamefont {Dutt},
  \citenamefont {Joshi}, \citenamefont {Ji}, \citenamefont {Cardenas},
  \citenamefont {Okawachi}, \citenamefont {Luke}, \citenamefont {Gaeta},\ and\
  \citenamefont {Lipson}}]{dutt_-chip_2018}%
  \BibitemOpen
  \bibfield  {author} {\bibinfo {author} {\bibfnamefont {A.}~\bibnamefont
  {Dutt}}, \bibinfo {author} {\bibfnamefont {C.}~\bibnamefont {Joshi}},
  \bibinfo {author} {\bibfnamefont {X.}~\bibnamefont {Ji}}, \bibinfo {author}
  {\bibfnamefont {J.}~\bibnamefont {Cardenas}}, \bibinfo {author}
  {\bibfnamefont {Y.}~\bibnamefont {Okawachi}}, \bibinfo {author}
  {\bibfnamefont {K.}~\bibnamefont {Luke}}, \bibinfo {author} {\bibfnamefont
  {A.~L.}\ \bibnamefont {Gaeta}}, \ and\ \bibinfo {author} {\bibfnamefont
  {M.}~\bibnamefont {Lipson}},\ }\href {\doibase 10.1126/sciadv.1701858}
  {\bibfield  {journal} {\bibinfo  {journal} {Science Advances}\ }\textbf
  {\bibinfo {volume} {4}},\ \bibinfo {pages} {e1701858} (\bibinfo {year}
  {2018})}\BibitemShut {NoStop}%
\bibitem [{\citenamefont {Trocha}\ \emph {et~al.}(2018)\citenamefont {Trocha},
  \citenamefont {Karpov}, \citenamefont {Ganin}, \citenamefont {Pfeiffer},
  \citenamefont {Kordts}, \citenamefont {Wolf}, \citenamefont {Krockenberger},
  \citenamefont {Marin-Palomo}, \citenamefont {Weimann}, \citenamefont
  {Randel}, \citenamefont {Freude}, \citenamefont {Kippenberg},\ and\
  \citenamefont {Koos}}]{trocha_ultrafast_2018}%
  \BibitemOpen
  \bibfield  {author} {\bibinfo {author} {\bibfnamefont {P.}~\bibnamefont
  {Trocha}}, \bibinfo {author} {\bibfnamefont {M.}~\bibnamefont {Karpov}},
  \bibinfo {author} {\bibfnamefont {D.}~\bibnamefont {Ganin}}, \bibinfo
  {author} {\bibfnamefont {M.~H.~P.}\ \bibnamefont {Pfeiffer}}, \bibinfo
  {author} {\bibfnamefont {A.}~\bibnamefont {Kordts}}, \bibinfo {author}
  {\bibfnamefont {S.}~\bibnamefont {Wolf}}, \bibinfo {author} {\bibfnamefont
  {J.}~\bibnamefont {Krockenberger}}, \bibinfo {author} {\bibfnamefont
  {P.}~\bibnamefont {Marin-Palomo}}, \bibinfo {author} {\bibfnamefont
  {C.}~\bibnamefont {Weimann}}, \bibinfo {author} {\bibfnamefont
  {S.}~\bibnamefont {Randel}}, \bibinfo {author} {\bibfnamefont
  {W.}~\bibnamefont {Freude}}, \bibinfo {author} {\bibfnamefont {T.~J.}\
  \bibnamefont {Kippenberg}}, \ and\ \bibinfo {author} {\bibfnamefont
  {C.}~\bibnamefont {Koos}},\ }\href {\doibase 10.1126/science.aao3924}
  {\bibfield  {journal} {\bibinfo  {journal} {Science}\ }\textbf {\bibinfo
  {volume} {359}},\ \bibinfo {pages} {887} (\bibinfo {year}
  {2018})}\BibitemShut {NoStop}%
\bibitem [{\citenamefont {Okawachi}\ \emph {et~al.}(2011)\citenamefont
  {Okawachi}, \citenamefont {Saha}, \citenamefont {Levy}, \citenamefont {Wen},
  \citenamefont {Lipson},\ and\ \citenamefont
  {Gaeta}}]{okawachi_octave-spanning_2011}%
  \BibitemOpen
  \bibfield  {author} {\bibinfo {author} {\bibfnamefont {Y.}~\bibnamefont
  {Okawachi}}, \bibinfo {author} {\bibfnamefont {K.}~\bibnamefont {Saha}},
  \bibinfo {author} {\bibfnamefont {J.~S.}\ \bibnamefont {Levy}}, \bibinfo
  {author} {\bibfnamefont {Y.~H.}\ \bibnamefont {Wen}}, \bibinfo {author}
  {\bibfnamefont {M.}~\bibnamefont {Lipson}}, \ and\ \bibinfo {author}
  {\bibfnamefont {A.~L.}\ \bibnamefont {Gaeta}},\ }\href {\doibase
  10.1364/OL.36.003398} {\bibfield  {journal} {\bibinfo  {journal} {Optics
  Letters}\ }\textbf {\bibinfo {volume} {36}},\ \bibinfo {pages} {3398}
  (\bibinfo {year} {2011})}\BibitemShut {NoStop}%
\bibitem [{\citenamefont {Griffith}\ \emph {et~al.}(2015)\citenamefont
  {Griffith}, \citenamefont {Lau}, \citenamefont {Cardenas}, \citenamefont
  {Okawachi}, \citenamefont {Mohanty}, \citenamefont {Fain}, \citenamefont
  {Lee}, \citenamefont {Yu}, \citenamefont {Phare}, \citenamefont {Poitras},
  \citenamefont {Gaeta},\ and\ \citenamefont
  {Lipson}}]{griffith_silicon-chip_2015}%
  \BibitemOpen
  \bibfield  {author} {\bibinfo {author} {\bibfnamefont {A.~G.}\ \bibnamefont
  {Griffith}}, \bibinfo {author} {\bibfnamefont {R.~K.~W.}\ \bibnamefont
  {Lau}}, \bibinfo {author} {\bibfnamefont {J.}~\bibnamefont {Cardenas}},
  \bibinfo {author} {\bibfnamefont {Y.}~\bibnamefont {Okawachi}}, \bibinfo
  {author} {\bibfnamefont {A.}~\bibnamefont {Mohanty}}, \bibinfo {author}
  {\bibfnamefont {R.}~\bibnamefont {Fain}}, \bibinfo {author} {\bibfnamefont
  {Y.~H.~D.}\ \bibnamefont {Lee}}, \bibinfo {author} {\bibfnamefont
  {M.}~\bibnamefont {Yu}}, \bibinfo {author} {\bibfnamefont {C.~T.}\
  \bibnamefont {Phare}}, \bibinfo {author} {\bibfnamefont {C.~B.}\ \bibnamefont
  {Poitras}}, \bibinfo {author} {\bibfnamefont {A.~L.}\ \bibnamefont {Gaeta}},
  \ and\ \bibinfo {author} {\bibfnamefont {M.}~\bibnamefont {Lipson}},\ }\href
  {\doibase 10.1038/ncomms7299} {\bibfield  {journal} {\bibinfo  {journal}
  {Nature Communications}\ }\textbf {\bibinfo {volume} {6}},\ \bibinfo {pages}
  {6299} (\bibinfo {year} {2015})}\BibitemShut {NoStop}%
\bibitem [{\citenamefont {Herr}\ \emph {et~al.}(2014)\citenamefont {Herr},
  \citenamefont {Brasch}, \citenamefont {Jost}, \citenamefont {Wang},
  \citenamefont {Kondratiev}, \citenamefont {Gorodetsky},\ and\ \citenamefont
  {Kippenberg}}]{herr_temporal_2014}%
  \BibitemOpen
  \bibfield  {author} {\bibinfo {author} {\bibfnamefont {T.}~\bibnamefont
  {Herr}}, \bibinfo {author} {\bibfnamefont {V.}~\bibnamefont {Brasch}},
  \bibinfo {author} {\bibfnamefont {J.~D.}\ \bibnamefont {Jost}}, \bibinfo
  {author} {\bibfnamefont {C.~Y.}\ \bibnamefont {Wang}}, \bibinfo {author}
  {\bibfnamefont {N.~M.}\ \bibnamefont {Kondratiev}}, \bibinfo {author}
  {\bibfnamefont {M.~L.}\ \bibnamefont {Gorodetsky}}, \ and\ \bibinfo {author}
  {\bibfnamefont {T.~J.}\ \bibnamefont {Kippenberg}},\ }\href {\doibase
  10.1038/nphoton.2013.343} {\bibfield  {journal} {\bibinfo  {journal} {Nature
  Photonics}\ }\textbf {\bibinfo {volume} {8}},\ \bibinfo {pages} {145}
  (\bibinfo {year} {2014})}\BibitemShut {NoStop}%
\bibitem [{\citenamefont {Hausmann}\ \emph {et~al.}(2014)\citenamefont
  {Hausmann}, \citenamefont {Bulu}, \citenamefont {Venkataraman}, \citenamefont
  {Deotare},\ and\ \citenamefont {Lončar}}]{hausmann_diamond_2014}%
  \BibitemOpen
  \bibfield  {author} {\bibinfo {author} {\bibfnamefont {B.~J.~M.}\
  \bibnamefont {Hausmann}}, \bibinfo {author} {\bibfnamefont {I.}~\bibnamefont
  {Bulu}}, \bibinfo {author} {\bibfnamefont {V.}~\bibnamefont {Venkataraman}},
  \bibinfo {author} {\bibfnamefont {P.}~\bibnamefont {Deotare}}, \ and\
  \bibinfo {author} {\bibfnamefont {M.}~\bibnamefont {Lončar}},\ }\href
  {\doibase 10.1038/nphoton.2014.72} {\bibfield  {journal} {\bibinfo  {journal}
  {Nature Photonics}\ }\textbf {\bibinfo {volume} {8}},\ \bibinfo {pages} {369}
  (\bibinfo {year} {2014})}\BibitemShut {NoStop}%
\bibitem [{\citenamefont {Jung}\ \emph
  {et~al.}(2014{\natexlab{a}})\citenamefont {Jung}, \citenamefont {Stoll},
  \citenamefont {Guo}, \citenamefont {Fischer},\ and\ \citenamefont
  {Tang}}]{jung_green_2014}%
  \BibitemOpen
  \bibfield  {author} {\bibinfo {author} {\bibfnamefont {H.}~\bibnamefont
  {Jung}}, \bibinfo {author} {\bibfnamefont {R.}~\bibnamefont {Stoll}},
  \bibinfo {author} {\bibfnamefont {X.}~\bibnamefont {Guo}}, \bibinfo {author}
  {\bibfnamefont {D.}~\bibnamefont {Fischer}}, \ and\ \bibinfo {author}
  {\bibfnamefont {H.~X.}\ \bibnamefont {Tang}},\ }\href {\doibase
  10.1364/OPTICA.1.000396} {\bibfield  {journal} {\bibinfo  {journal} {Optica}\
  }\textbf {\bibinfo {volume} {1}},\ \bibinfo {pages} {396} (\bibinfo {year}
  {2014}{\natexlab{a}})}\BibitemShut {NoStop}%
\bibitem [{\citenamefont {Pu}\ \emph {et~al.}(2016)\citenamefont {Pu},
  \citenamefont {Ottaviano}, \citenamefont {Semenova},\ and\ \citenamefont
  {Yvind}}]{pu_efficient_2016}%
  \BibitemOpen
  \bibfield  {author} {\bibinfo {author} {\bibfnamefont {M.}~\bibnamefont
  {Pu}}, \bibinfo {author} {\bibfnamefont {L.}~\bibnamefont {Ottaviano}},
  \bibinfo {author} {\bibfnamefont {E.}~\bibnamefont {Semenova}}, \ and\
  \bibinfo {author} {\bibfnamefont {K.}~\bibnamefont {Yvind}},\ }\href
  {\doibase 10.1364/OPTICA.3.000823} {\bibfield  {journal} {\bibinfo  {journal}
  {Optica}\ }\textbf {\bibinfo {volume} {3}},\ \bibinfo {pages} {823} (\bibinfo
  {year} {2016})}\BibitemShut {NoStop}%
\bibitem [{\citenamefont {Reed}\ \emph {et~al.}(2010)\citenamefont {Reed},
  \citenamefont {Mashanovich}, \citenamefont {Gardes},\ and\ \citenamefont
  {Thomson}}]{reed_silicon_2010}%
  \BibitemOpen
  \bibfield  {author} {\bibinfo {author} {\bibfnamefont {G.~T.}\ \bibnamefont
  {Reed}}, \bibinfo {author} {\bibfnamefont {G.}~\bibnamefont {Mashanovich}},
  \bibinfo {author} {\bibfnamefont {F.~Y.}\ \bibnamefont {Gardes}}, \ and\
  \bibinfo {author} {\bibfnamefont {D.~J.}\ \bibnamefont {Thomson}},\ }\href
  {\doibase 10.1038/nphoton.2010.179} {\bibfield  {journal} {\bibinfo
  {journal} {Nature Photonics}\ }\textbf {\bibinfo {volume} {4}},\ \bibinfo
  {pages} {518} (\bibinfo {year} {2010})}\BibitemShut {NoStop}%
\bibitem [{\citenamefont {Nikogosyan}(2005)}]{nikogosyan_nonlinear_2005}%
  \BibitemOpen
  \bibfield  {author} {\bibinfo {author} {\bibfnamefont {D.~N.}\ \bibnamefont
  {Nikogosyan}},\ }\href {//www.springer.com/us/book/9780387220222} {\emph
  {\bibinfo {title} {Nonlinear {Optical} {Crystals}: {A} {Complete}
  {Survey}}}}\ (\bibinfo  {publisher} {Springer-Verlag},\ \bibinfo {address}
  {New York},\ \bibinfo {year} {2005})\BibitemShut {NoStop}%
\bibitem [{\citenamefont {Miller}\ \emph {et~al.}(2015)\citenamefont {Miller},
  \citenamefont {Okawachi}, \citenamefont {Ramelow}, \citenamefont {Luke},
  \citenamefont {Dutt}, \citenamefont {Farsi}, \citenamefont {Gaeta},\ and\
  \citenamefont {Lipson}}]{miller_tunable_2015}%
  \BibitemOpen
  \bibfield  {author} {\bibinfo {author} {\bibfnamefont {S.~A.}\ \bibnamefont
  {Miller}}, \bibinfo {author} {\bibfnamefont {Y.}~\bibnamefont {Okawachi}},
  \bibinfo {author} {\bibfnamefont {S.}~\bibnamefont {Ramelow}}, \bibinfo
  {author} {\bibfnamefont {K.}~\bibnamefont {Luke}}, \bibinfo {author}
  {\bibfnamefont {A.}~\bibnamefont {Dutt}}, \bibinfo {author} {\bibfnamefont
  {A.}~\bibnamefont {Farsi}}, \bibinfo {author} {\bibfnamefont {A.~L.}\
  \bibnamefont {Gaeta}}, \ and\ \bibinfo {author} {\bibfnamefont
  {M.}~\bibnamefont {Lipson}},\ }\href {\doibase 10.1364/OE.23.021527}
  {\bibfield  {journal} {\bibinfo  {journal} {Optics Express}\ }\textbf
  {\bibinfo {volume} {23}},\ \bibinfo {pages} {21527} (\bibinfo {year}
  {2015})}\BibitemShut {NoStop}%
\bibitem [{\citenamefont {Jung}\ \emph
  {et~al.}(2014{\natexlab{b}})\citenamefont {Jung}, \citenamefont {Fong},
  \citenamefont {Xiong},\ and\ \citenamefont {Tang}}]{jung_electrical_2014}%
  \BibitemOpen
  \bibfield  {author} {\bibinfo {author} {\bibfnamefont {H.}~\bibnamefont
  {Jung}}, \bibinfo {author} {\bibfnamefont {K.~Y.}\ \bibnamefont {Fong}},
  \bibinfo {author} {\bibfnamefont {C.}~\bibnamefont {Xiong}}, \ and\ \bibinfo
  {author} {\bibfnamefont {H.~X.}\ \bibnamefont {Tang}},\ }\href {\doibase
  10.1364/OL.39.000084} {\bibfield  {journal} {\bibinfo  {journal} {Optics
  Letters}\ }\textbf {\bibinfo {volume} {39}},\ \bibinfo {pages} {84} (\bibinfo
  {year} {2014}{\natexlab{b}})}\BibitemShut {NoStop}%
\bibitem [{\citenamefont {Spencer}\ \emph {et~al.}(2018)\citenamefont
  {Spencer}, \citenamefont {Drake}, \citenamefont {Briles}, \citenamefont
  {Stone}, \citenamefont {Sinclair}, \citenamefont {Fredrick}, \citenamefont
  {Li}, \citenamefont {Westly}, \citenamefont {Ilic}, \citenamefont
  {Bluestone}, \citenamefont {Volet}, \citenamefont {Komljenovic},
  \citenamefont {Chang}, \citenamefont {Lee}, \citenamefont {Oh}, \citenamefont
  {Suh}, \citenamefont {Yang}, \citenamefont {Pfeiffer}, \citenamefont
  {Kippenberg}, \citenamefont {Norberg}, \citenamefont {Theogarajan},
  \citenamefont {Vahala}, \citenamefont {Newbury}, \citenamefont {Srinivasan},
  \citenamefont {Bowers}, \citenamefont {Diddams},\ and\ \citenamefont
  {Papp}}]{spencer_optical-frequency_2018}%
  \BibitemOpen
  \bibfield  {author} {\bibinfo {author} {\bibfnamefont {D.~T.}\ \bibnamefont
  {Spencer}}, \bibinfo {author} {\bibfnamefont {T.}~\bibnamefont {Drake}},
  \bibinfo {author} {\bibfnamefont {T.~C.}\ \bibnamefont {Briles}}, \bibinfo
  {author} {\bibfnamefont {J.}~\bibnamefont {Stone}}, \bibinfo {author}
  {\bibfnamefont {L.~C.}\ \bibnamefont {Sinclair}}, \bibinfo {author}
  {\bibfnamefont {C.}~\bibnamefont {Fredrick}}, \bibinfo {author}
  {\bibfnamefont {Q.}~\bibnamefont {Li}}, \bibinfo {author} {\bibfnamefont
  {D.}~\bibnamefont {Westly}}, \bibinfo {author} {\bibfnamefont {B.~R.}\
  \bibnamefont {Ilic}}, \bibinfo {author} {\bibfnamefont {A.}~\bibnamefont
  {Bluestone}}, \bibinfo {author} {\bibfnamefont {N.}~\bibnamefont {Volet}},
  \bibinfo {author} {\bibfnamefont {T.}~\bibnamefont {Komljenovic}}, \bibinfo
  {author} {\bibfnamefont {L.}~\bibnamefont {Chang}}, \bibinfo {author}
  {\bibfnamefont {S.~H.}\ \bibnamefont {Lee}}, \bibinfo {author} {\bibfnamefont
  {D.~Y.}\ \bibnamefont {Oh}}, \bibinfo {author} {\bibfnamefont {M.-G.}\
  \bibnamefont {Suh}}, \bibinfo {author} {\bibfnamefont {K.~Y.}\ \bibnamefont
  {Yang}}, \bibinfo {author} {\bibfnamefont {M.~H.~P.}\ \bibnamefont
  {Pfeiffer}}, \bibinfo {author} {\bibfnamefont {T.~J.}\ \bibnamefont
  {Kippenberg}}, \bibinfo {author} {\bibfnamefont {E.}~\bibnamefont {Norberg}},
  \bibinfo {author} {\bibfnamefont {L.}~\bibnamefont {Theogarajan}}, \bibinfo
  {author} {\bibfnamefont {K.}~\bibnamefont {Vahala}}, \bibinfo {author}
  {\bibfnamefont {N.~R.}\ \bibnamefont {Newbury}}, \bibinfo {author}
  {\bibfnamefont {K.}~\bibnamefont {Srinivasan}}, \bibinfo {author}
  {\bibfnamefont {J.~E.}\ \bibnamefont {Bowers}}, \bibinfo {author}
  {\bibfnamefont {S.~A.}\ \bibnamefont {Diddams}}, \ and\ \bibinfo {author}
  {\bibfnamefont {S.~B.}\ \bibnamefont {Papp}},\ }\href {\doibase
  10.1038/s41586-018-0065-7} {\bibfield  {journal} {\bibinfo  {journal}
  {Nature}\ }\textbf {\bibinfo {volume} {557}},\ \bibinfo {pages} {81}
  (\bibinfo {year} {2018})}\BibitemShut {NoStop}%
\bibitem [{\citenamefont {DeSalvo}\ \emph {et~al.}(1996)\citenamefont
  {DeSalvo}, \citenamefont {Said}, \citenamefont {Hagan}, \citenamefont
  {Stryland},\ and\ \citenamefont {Sheik-Bahae}}]{desalvo_infrared_1996}%
  \BibitemOpen
  \bibfield  {author} {\bibinfo {author} {\bibfnamefont {R.}~\bibnamefont
  {DeSalvo}}, \bibinfo {author} {\bibfnamefont {A.~A.}\ \bibnamefont {Said}},
  \bibinfo {author} {\bibfnamefont {D.~J.}\ \bibnamefont {Hagan}}, \bibinfo
  {author} {\bibfnamefont {E.~W.~V.}\ \bibnamefont {Stryland}}, \ and\ \bibinfo
  {author} {\bibfnamefont {M.}~\bibnamefont {Sheik-Bahae}},\ }\href {\doibase
  10.1109/3.511545} {\bibfield  {journal} {\bibinfo  {journal} {IEEE Journal of
  Quantum Electronics}\ }\textbf {\bibinfo {volume} {32}},\ \bibinfo {pages}
  {1324} (\bibinfo {year} {1996})}\BibitemShut {NoStop}%
\bibitem [{\citenamefont {Ilchenko}\ \emph {et~al.}(2004)\citenamefont
  {Ilchenko}, \citenamefont {Savchenkov}, \citenamefont {Matsko},\ and\
  \citenamefont {Maleki}}]{ilchenko_nonlinear_2004}%
  \BibitemOpen
  \bibfield  {author} {\bibinfo {author} {\bibfnamefont {V.~S.}\ \bibnamefont
  {Ilchenko}}, \bibinfo {author} {\bibfnamefont {A.~A.}\ \bibnamefont
  {Savchenkov}}, \bibinfo {author} {\bibfnamefont {A.~B.}\ \bibnamefont
  {Matsko}}, \ and\ \bibinfo {author} {\bibfnamefont {L.}~\bibnamefont
  {Maleki}},\ }\href {\doibase 10.1103/PhysRevLett.92.043903} {\bibfield
  {journal} {\bibinfo  {journal} {Physical Review Letters}\ }\textbf {\bibinfo
  {volume} {92}},\ \bibinfo {pages} {043903} (\bibinfo {year}
  {2004})}\BibitemShut {NoStop}%
\bibitem [{\citenamefont {Wang}\ \emph {et~al.}(2018)\citenamefont {Wang},
  \citenamefont {Zhang}, \citenamefont {Stern}, \citenamefont {Lipson},\ and\
  \citenamefont {Lončar}}]{wang_nanophotonic_2018}%
  \BibitemOpen
  \bibfield  {author} {\bibinfo {author} {\bibfnamefont {C.}~\bibnamefont
  {Wang}}, \bibinfo {author} {\bibfnamefont {M.}~\bibnamefont {Zhang}},
  \bibinfo {author} {\bibfnamefont {B.}~\bibnamefont {Stern}}, \bibinfo
  {author} {\bibfnamefont {M.}~\bibnamefont {Lipson}}, \ and\ \bibinfo {author}
  {\bibfnamefont {M.}~\bibnamefont {Lončar}},\ }\href {\doibase
  10.1364/OE.26.001547} {\bibfield  {journal} {\bibinfo  {journal} {Optics
  Express}\ }\textbf {\bibinfo {volume} {26}},\ \bibinfo {pages} {1547}
  (\bibinfo {year} {2018})}\BibitemShut {NoStop}%
\bibitem [{\citenamefont {Zhang}\ \emph {et~al.}(2017)\citenamefont {Zhang},
  \citenamefont {Wang}, \citenamefont {Cheng}, \citenamefont {Shams-Ansari},\
  and\ \citenamefont {Lončar}}]{zhang_monolithic_2017}%
  \BibitemOpen
  \bibfield  {author} {\bibinfo {author} {\bibfnamefont {M.}~\bibnamefont
  {Zhang}}, \bibinfo {author} {\bibfnamefont {C.}~\bibnamefont {Wang}},
  \bibinfo {author} {\bibfnamefont {R.}~\bibnamefont {Cheng}}, \bibinfo
  {author} {\bibfnamefont {A.}~\bibnamefont {Shams-Ansari}}, \ and\ \bibinfo
  {author} {\bibfnamefont {M.}~\bibnamefont {Lončar}},\ }\href {\doibase
  10.1364/OPTICA.4.001536} {\bibfield  {journal} {\bibinfo  {journal} {Optica}\
  }\textbf {\bibinfo {volume} {4}},\ \bibinfo {pages} {1536} (\bibinfo {year}
  {2017})}\BibitemShut {NoStop}%
\bibitem [{\citenamefont {He}\ \emph {et~al.}(2018)\citenamefont {He},
  \citenamefont {Liang}, \citenamefont {Luo}, \citenamefont {Lin},\ and\
  \citenamefont {Lin}}]{he_dispersion-engineered_2018}%
  \BibitemOpen
  \bibfield  {author} {\bibinfo {author} {\bibfnamefont {Y.}~\bibnamefont
  {He}}, \bibinfo {author} {\bibfnamefont {H.}~\bibnamefont {Liang}}, \bibinfo
  {author} {\bibfnamefont {R.}~\bibnamefont {Luo}}, \bibinfo {author}
  {\bibfnamefont {Q.}~\bibnamefont {Lin}}, \ and\ \bibinfo {author}
  {\bibfnamefont {Q.}~\bibnamefont {Lin}},\ }in\ \href {\doibase
  10.1364/CLEO_AT.2018.JW2A.64} {\emph {\bibinfo {booktitle} {Conference on
  {Lasers} and {Electro}-{Optics} (2018), paper {JW}2A.64}}}\ (\bibinfo
  {publisher} {Optical Society of America},\ \bibinfo {year} {2018})\ p.\
  \bibinfo {pages} {JW2A.64}\BibitemShut {NoStop}%
\bibitem [{\citenamefont {Wang}\ \emph {et~al.}(2014)\citenamefont {Wang},
  \citenamefont {Burek}, \citenamefont {Lin}, \citenamefont {Atikian},
  \citenamefont {Venkataraman}, \citenamefont {Huang}, \citenamefont {Stark},\
  and\ \citenamefont {Lončar}}]{wang_integrated_2014}%
  \BibitemOpen
  \bibfield  {author} {\bibinfo {author} {\bibfnamefont {C.}~\bibnamefont
  {Wang}}, \bibinfo {author} {\bibfnamefont {M.~J.}\ \bibnamefont {Burek}},
  \bibinfo {author} {\bibfnamefont {Z.}~\bibnamefont {Lin}}, \bibinfo {author}
  {\bibfnamefont {H.~A.}\ \bibnamefont {Atikian}}, \bibinfo {author}
  {\bibfnamefont {V.}~\bibnamefont {Venkataraman}}, \bibinfo {author}
  {\bibfnamefont {I.-C.}\ \bibnamefont {Huang}}, \bibinfo {author}
  {\bibfnamefont {P.}~\bibnamefont {Stark}}, \ and\ \bibinfo {author}
  {\bibfnamefont {M.}~\bibnamefont {Lončar}},\ }\href {\doibase
  10.1364/OE.22.030924} {\bibfield  {journal} {\bibinfo  {journal} {Optics
  Express}\ }\textbf {\bibinfo {volume} {22}},\ \bibinfo {pages} {30924}
  (\bibinfo {year} {2014})}\BibitemShut {NoStop}%
\bibitem [{\citenamefont {Wolf}\ \emph {et~al.}(2017)\citenamefont {Wolf},
  \citenamefont {Breunig}, \citenamefont {Zappe},\ and\ \citenamefont
  {Buse}}]{wolf_cascaded_2017}%
  \BibitemOpen
  \bibfield  {author} {\bibinfo {author} {\bibfnamefont {R.}~\bibnamefont
  {Wolf}}, \bibinfo {author} {\bibfnamefont {I.}~\bibnamefont {Breunig}},
  \bibinfo {author} {\bibfnamefont {H.}~\bibnamefont {Zappe}}, \ and\ \bibinfo
  {author} {\bibfnamefont {K.}~\bibnamefont {Buse}},\ }\href {\doibase
  10.1364/OE.25.029927} {\bibfield  {journal} {\bibinfo  {journal} {Optics
  Express}\ }\textbf {\bibinfo {volume} {25}},\ \bibinfo {pages} {29927}
  (\bibinfo {year} {2017})}\BibitemShut {NoStop}%
\bibitem [{\citenamefont {Wang}\ \emph {et~al.}(2015)\citenamefont {Wang},
  \citenamefont {Bo}, \citenamefont {Wan}, \citenamefont {Li}, \citenamefont
  {Gao}, \citenamefont {Li}, \citenamefont {Zhang},\ and\ \citenamefont
  {Xu}}]{wang_high-q_2015}%
  \BibitemOpen
  \bibfield  {author} {\bibinfo {author} {\bibfnamefont {J.}~\bibnamefont
  {Wang}}, \bibinfo {author} {\bibfnamefont {F.}~\bibnamefont {Bo}}, \bibinfo
  {author} {\bibfnamefont {S.}~\bibnamefont {Wan}}, \bibinfo {author}
  {\bibfnamefont {W.}~\bibnamefont {Li}}, \bibinfo {author} {\bibfnamefont
  {F.}~\bibnamefont {Gao}}, \bibinfo {author} {\bibfnamefont {J.}~\bibnamefont
  {Li}}, \bibinfo {author} {\bibfnamefont {G.}~\bibnamefont {Zhang}}, \ and\
  \bibinfo {author} {\bibfnamefont {J.}~\bibnamefont {Xu}},\ }\href {\doibase
  10.1364/OE.23.023072} {\bibfield  {journal} {\bibinfo  {journal} {Optics
  Express}\ }\textbf {\bibinfo {volume} {23}},\ \bibinfo {pages} {23072}
  (\bibinfo {year} {2015})}\BibitemShut {NoStop}%
\bibitem [{\citenamefont {Liang}\ \emph {et~al.}(2017)\citenamefont {Liang},
  \citenamefont {Luo}, \citenamefont {He}, \citenamefont {Jiang},\ and\
  \citenamefont {Lin}}]{liang_high-quality_2017}%
  \BibitemOpen
  \bibfield  {author} {\bibinfo {author} {\bibfnamefont {H.}~\bibnamefont
  {Liang}}, \bibinfo {author} {\bibfnamefont {R.}~\bibnamefont {Luo}}, \bibinfo
  {author} {\bibfnamefont {Y.}~\bibnamefont {He}}, \bibinfo {author}
  {\bibfnamefont {H.}~\bibnamefont {Jiang}}, \ and\ \bibinfo {author}
  {\bibfnamefont {Q.}~\bibnamefont {Lin}},\ }\href {\doibase
  10.1364/OPTICA.4.001251} {\bibfield  {journal} {\bibinfo  {journal} {Optica}\
  }\textbf {\bibinfo {volume} {4}},\ \bibinfo {pages} {1251} (\bibinfo {year}
  {2017})}\BibitemShut {NoStop}%
\bibitem [{\citenamefont {Zhang}\ \emph {et~al.}(2018)\citenamefont {Zhang},
  \citenamefont {Wang}, \citenamefont {Chen}, \citenamefont {Bertrand},
  \citenamefont {Bertrand}, \citenamefont {Shams-Ansari}, \citenamefont
  {Chandrasekhar}, \citenamefont {Winzer},\ and\ \citenamefont
  {Lončar}}]{zhang_ultra-high_2018}%
  \BibitemOpen
  \bibfield  {author} {\bibinfo {author} {\bibfnamefont {M.}~\bibnamefont
  {Zhang}}, \bibinfo {author} {\bibfnamefont {C.}~\bibnamefont {Wang}},
  \bibinfo {author} {\bibfnamefont {X.}~\bibnamefont {Chen}}, \bibinfo {author}
  {\bibfnamefont {M.}~\bibnamefont {Bertrand}}, \bibinfo {author}
  {\bibfnamefont {M.}~\bibnamefont {Bertrand}}, \bibinfo {author}
  {\bibfnamefont {A.}~\bibnamefont {Shams-Ansari}}, \bibinfo {author}
  {\bibfnamefont {S.}~\bibnamefont {Chandrasekhar}}, \bibinfo {author}
  {\bibfnamefont {P.}~\bibnamefont {Winzer}}, \ and\ \bibinfo {author}
  {\bibfnamefont {M.}~\bibnamefont {Lončar}},\ }in\ \href {\doibase
  10.1364/OFC.2018.Th4A.5} {\emph {\bibinfo {booktitle} {Optical {Fiber}
  {Communication} {Conference} {Postdeadline} {Papers} (2018), paper
  {Th}4A.5}}}\ (\bibinfo  {publisher} {Optical Society of America},\ \bibinfo
  {year} {2018})\ p.\ \bibinfo {pages} {Th4A.5}\BibitemShut {NoStop}%
\bibitem [{\citenamefont {Phillips}\ \emph {et~al.}(2011)\citenamefont
  {Phillips}, \citenamefont {Langrock}, \citenamefont {Pelc}, \citenamefont
  {Fejer}, \citenamefont {Hartl},\ and\ \citenamefont
  {Fermann}}]{phillips_supercontinuum_2011}%
  \BibitemOpen
  \bibfield  {author} {\bibinfo {author} {\bibfnamefont {C.~R.}\ \bibnamefont
  {Phillips}}, \bibinfo {author} {\bibfnamefont {C.}~\bibnamefont {Langrock}},
  \bibinfo {author} {\bibfnamefont {J.~S.}\ \bibnamefont {Pelc}}, \bibinfo
  {author} {\bibfnamefont {M.~M.}\ \bibnamefont {Fejer}}, \bibinfo {author}
  {\bibfnamefont {I.}~\bibnamefont {Hartl}}, \ and\ \bibinfo {author}
  {\bibfnamefont {M.~E.}\ \bibnamefont {Fermann}},\ }\href {\doibase
  10.1364/OE.19.018754} {\bibfield  {journal} {\bibinfo  {journal} {Optics
  Express}\ }\textbf {\bibinfo {volume} {19}},\ \bibinfo {pages} {18754}
  (\bibinfo {year} {2011})}\BibitemShut {NoStop}%
\end{thebibliography}%

\pagebreak
\onecolumngrid

\begin{center}
	\textbf{\large Supplemental Information}
\end{center}

\setcounter{equation}{0}
\setcounter{figure}{0}
\setcounter{table}{0}
\setcounter{page}{1}
\makeatletter
\renewcommand{\theequation}{S\arabic{equation}}
\renewcommand{\thefigure}{S\arabic{figure}}
\renewcommand{\bibnumfmt}[1]{[S#1]}
\renewcommand{\citenumfont}[1]{S#1}
\section{Device design and simulation}
Waveguide dispersion diagrams and mode profiles are numerically calculated using a commercial Finite Difference Eigenmode (FDE) solver (Lumerical, Mode Solutions). Numerical simulation shows that, for the current device layer thickness of 600 nm, air cladding is necessary for anomalous dispersions. For the filter ring, however, a SiO$_2$ cladding gives rise to a better electro-optic tuning efficiency \cite{wang_nanophotonic_2018}. Therefore in the final chip, the SiO$_2$ cladding in the comb generator area is intentionally removed, while the rest of the chip, including the filter ring, is cladded (Fig. \ref{fig1}). The filter tuning efficiency of 2.4 pm/V is lower than our previous results \cite{wang_nanophotonic_2018} since only one arm of the ring resonator is modulated. 

\section{Device fabrication}
Devices are fabricated from a commercial x-cut LN-on-insulator (LNOI) wafer (NANOLN) with a 600-nm device layer thickness. Electron-beam lithography (EBL) and Ar+-based reactive ion etching (RIE) are used to create optical waveguides and microring resonators in the LN film, using a similar process as described in our previous work \cite{zhang_monolithic_2017}. A 1.5-$\mu$m-thick PMMA EBL resist is spun coated and exposed using a second EBL with alignment, to produce the microelectrodes of the filter ring via a lift-off process. The structures are then cladded by an 800-nm-thick SiO$_2$ layer using plasma-enhanced chemical vapor deposition (PECVD). The oxide cladding in the comb generation areas are then removed through a photolithography step followed by hydrofluoric acid (HF) wet etching to realize air-cladded devices with the required anomalous dispersions. Finally, the chip edges are diced and polished to improve the fiber-chip coupling. 

\section{Characterization of the comb generation, filtering and modulation}
For frequency comb characterization, continuous-wave (CW) light from a tuneable telecom laser (Santec TSL-510) is amplified using an erbium-doped fiber amplifier (EDFA, Amonics). A 3-paddle fibre polarization controller is used to control the polarization of input light. Tapered lensed fibres are used to couple light into and out from the waveguide facets of the LN chip. The output light is sent into an optical spectrum analyzer (OSA, Yokogawa) for analysis. For filter control and manipulation, TE polarized modes are used to exploit the highest electro-optic tuning efficiency. DC signals from a voltage supply (Keithley) and AC signals from an arbitrary waveform generator (AWG, Tektronix 70001A) are combined using a bias T, before being sent to the filter electrodes using a high-speed ground-signal (GS) probe (GGB Industries). The output optical signal from the drop port is sent to a 12-GHz photodetector (Newport 1544A), and analysed using a 1-GHz real-time oscilloscope (Tektronix).
 
\section{Optical parametric oscillation threshold}
The threshold power of the OPO process can be estimated as \cite{kippenberg_kerr-nonlinearity_2004}

\begin{equation}
\label{eq:th}
P_{\textrm{th}}\approx 1.54\frac{\pi}{2}\frac{Q_c}{2 Q_L}\frac{n_{\textrm{eff}^2} V}{n_2 \lambda_p Q_L^2}
\end{equation}

where $Q_C$ and $Q_L$ are the coupling and loaded $Q$ factors of the resonator, $n_\textrm{eff}$ is the effective refractive index of the LN waveguide, $n_2$ is the nonlinear refractive index, $\lambda_p$ is the pump wavelength and $V$ is the resonator mode volume. 

From the measured $Q_L = 6.6 \times 10^5$ and the optical transmission depth $T = 4.6\%$ for TE mode (Fig. 1b), we estimate the intrinsic quality factor $Q_i=(2Q_L)/(1+\sqrt{T})=1.1\times 10^6$ assuming the device is under-coupled. Since $Q_L^{-1} = Q_C^{-1} + Q_i^{-1}$, the coupling $Q$ is calculated to be $Q_C = 1.7 \times 10^6$. We numerically calculate the effective index of our LN waveguide to be $n_\textrm{eff}$ = 1.91 and the mode area $A = 0.875 \mu$m$^2$. The resonator mode volume is $V = 2\pi RA = 440 \mu$m$^3$, where $R = 80 \mu$m is the radius of the ring resonator. The nonlinear refractive index $n_2 = 0.91 \times 10^{-15}$ cm$^2$/W. From these parameters we calculate the optical parametric oscillation threshold to be $P_{\textrm{th}} = 80$ mW.

\section{Filter transmission spectra}
\begin{figure*}
	\centering
	\includegraphics[angle=0,width=\textwidth]{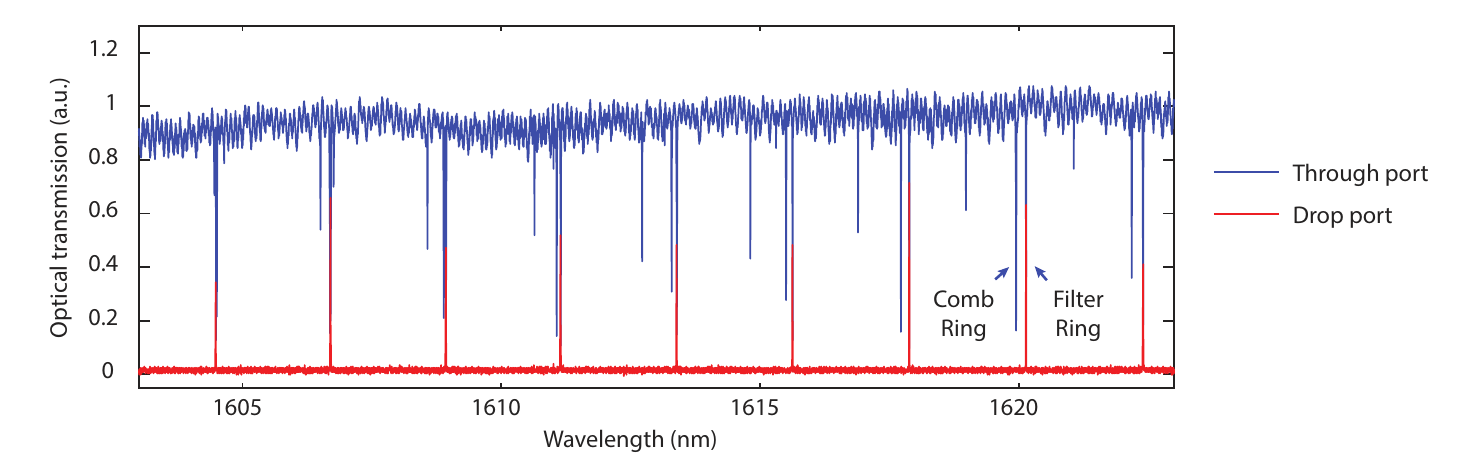}
	
	\caption{\label{figs1} \textbf{Optical transmission spectra of the device at both through and drop ports of the filter.} The resonances of the comb generator and the filter are aligned near 1606 nm, and have a $1.2\%$ mismatch in the free spectral range.}
\end{figure*}

The filter ring resonator is designed to have a free spectral range (FSR) that is $1.2\%$ larger than the comb-generating resonator. Supplementary Figure \ref{figs1} shows the optical transmission spectra at both through and drop ports of the filter, measured using a tunable telecom laser. The transmission dips at the through port correspond to resonances of both ring resonators, while the transmission peaks at the drop port correspond to the filter passband. The spectra show that the two ring resonances are aligned near 1606 nm, and have increasing mismatch when the wavelength is away from 1606 nm. 

\section{Filter transfer function}
\begin{figure*}
	\centering
	\includegraphics[angle=0,width=0.7\textwidth]{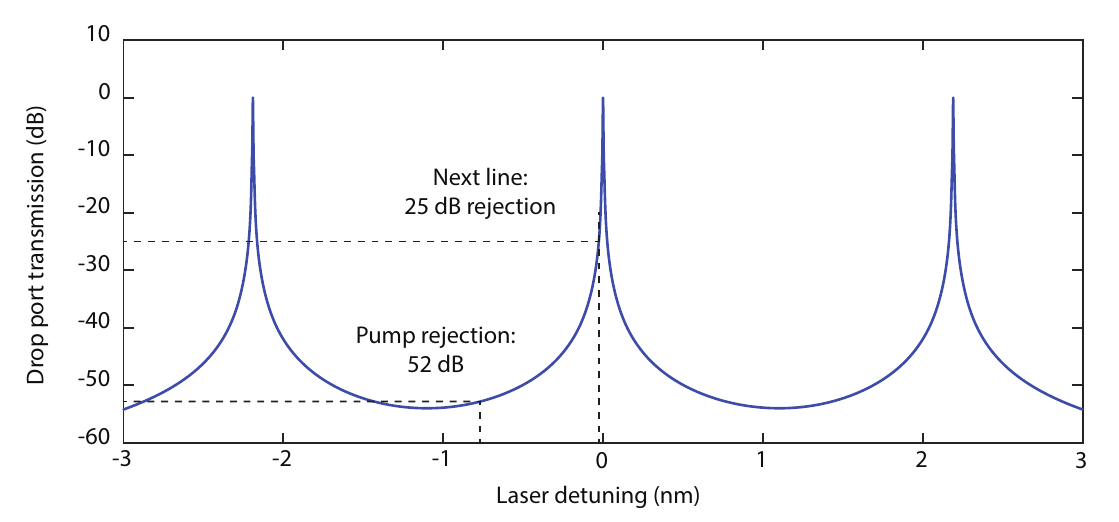}
	
	\caption{\label{figs2} \textbf{Numerically calculated filter transfer function.} The calculated suppression ratios for the next line and the pump are 25 dB and 52 dB respectively.}
\end{figure*}

Supplementary Figure \ref{figs2} shows the theoretical transfer function of the filter, which has a FSR of 2.19 nm and a linewidth of 3 pm. The calculated suppression ratio for the next comb line, which has a 27-pm resonance mismatch, is 25 dB. The calculated suppression ratio for the pump light (resonance mismatch by 730 pm) is 52 dB.

\end{document}